**Potential-tuning molecular dynamics studies of fusion, and the question of ideal glassformers: (I) The Gay-Berne model.**

Vitaliy Kapko, Dmitry V. Matyushov and C. Austen Angell,
Department of Chemistry and Biochemistry,
Arizona State University, Tempe, AZ85287-1604.

**Abstract**
**The ability of some liquids to vitrify during supercooling is usually seen as a consequence of the rates of crystal nucleation (and/or crystal growth) becoming small[1]- thus a matter of kinetics. However there is evidence, dating back to the empirics of coal briquetting for maximum trucking efficiency[2] that some object shapes find little advantage in self-assembly to ordered structures - meaning random packings prevail. Noting that key studies of non-spherical object packing have never been followed from hard ellipsoids[3,4] or spherocylinders[5] (diatomics excepted[6]) into the world of molecules with attractive forces, we have made a molecular dynamics MD study of crystal melting and glass formation on the Gay-Berne (G-B) model of ellipsoidal objects[7] across the aspect ratio range of the hard ellipsoid studies. Here we report that, in the aspect ratio range of maximum ellipsoid packing efficiency, various G-B crystalline states, that cannot be obtained directly from the liquid, disorder spontaneously near 0 K and transform to liquids without any detectable enthalpy of fusion. Without claiming to have proved the existence of single component examples, we use the present observations, together with our knowledge of non-ideal mixing effects, to discuss the probable existence of "ideal glassformers" – single or multicomponent liquids that vitrify before ever becoming metastable with respect to crystals. The existence of crystal-free routes to the glassy state removes any precrystalline fluctuation perspective from the "glass problem"[8]. Unexpectedly we find that liquids with aspect ratios in the "crystallophobic" range also behave in an unusual (non-hysteritic) way during temperature cycling through the glass transition. We link this to the highly volume fraction-sensitive ("fragile") behavior observed in recent hard dumbbell studies at similar length/diameter ratios[9].**

It is generally thought that, to understand glass formation, the kinetics of crystal nucleation and growth must be understood in detail. This is because of the experience that, given sufficient opportunity, common glassformers (some atactic polymers excepted) seem to convert at least partly into crystalline materials. The feeling that glasses must always be metastable with respect to crystallization is also encouraged by the popular "2/3" rule which holds that if a liquid fails to crystallize on cooling it will become a brittle glass at 2/3 of its melting point. We have argued elsewhere[10] that this rule is a tautology, originating in our inability to obtain data on systems that do not more-or-less satisfy the rule. Clearly a liquid that never crystallizes cannot be tested for adherence to the rule[11]. Likewise, liquids with melting points near their boiling points will not be vitrifiable so cannot be tested. Indeed, for the majority of metallic glasses (until the recent development of "bulk" variants that do not require enormous cooling rates), crystallization occurred during reheating before a glass transition could ever be observed.

In a recent study of protic ionic liquids, most of which are glassforming, a broad scatter of $T_g/T_m$ values was found about an average value of ~2/3. The cut-off for crystallizing ability seemed to be about $T_g/T_m = 0.85$[10]. Likewise, in the study of molecular liquids of interest to the pharmaceutical industry a wide distribution of $T_g/T_m$ values has been recorded, with examples as high as $0.86$[12]. This has provoked our interest in the possibility of cases where $T_g/T_m > 1.0$. Some non-polymeric cases exist in practice, e.g. that of the eutectic in the binary system $H_2Cr_2O_7 + H_2O$ near the "failed crystal" hexahydrate composition[13], but are rare, or rarely studied: (in solutions, the ideal glassformer criterion must be understood in terms of chemical potentials in solution remaining below those of any crystals). Recent theoretical work has revealed 2D systems with amorphous ground states[14].

A particular reason for being interested in such cases is the way in which it changes our viewpoint on the understanding of glassforming ability[11]. Clearly, for a substance that vitrifies before it becomes metastable, the rate of nucleation of a crystalline phase becomes quite irrelevant. Instead of asking why some liquids nucleate slowly it forces us to ask a quite different question, namely, what is it that renders some substances incapable of forming a 3D-ordered lattice (or a combination of them in the case of multicomponent systems) that can compete, in chemical potential, with its disordered cousin (or their disordered solution, in the case of multicomponent systems). An additional, and powerful, reason for being interested in such a problem is that it becomes a problem accessible to fruitful study by computer simulation. Molecular dynamics may be inadequate to study crystal nucleation in a liquid with a relaxation time of 100 sec, but it is ideally suited to measuring the heat of fusion of a preformed crystal. Thus Molinero et al, using "potential-tuning" molecular dynamics, PTMD, were able to report, recently, the melting points and heats of fusion for a range of monatomic systems related to the tetrahedrally coordinated case of silicon, some of which were so stable, kinetically, in the liquid state, that they could never be observed to recrystallize once they had been heated past melting[15].

Here we will use this same approach (Supplementary Information, SI-1) to demonstrate the existence of much more extreme behavior in the case of systems of continuously variable shape. We first confirm (SI-2) that the melting point measuring procedure adopted by Molinero et al for their study[15] can, in short time, give results of good quantitative accuracy. Then we will describe the results of applying the same PTMD strategy for fusion studies to a quite different type of system, one that can be investigated with pairwise additive potentials. We apply it to the study of liquids of short ellipsoidal molecules which, in the $\alpha = 1$ limit of diminishing aspect ratio $\alpha$, become the standard Lennard-Jones (LJ) liquid.

To do this we use the Gay-Berne (G-B) model[7] developed for study of liquid crystals[16,17]. The model, and our simulation methods, are detailed in SI-1. For a 1000 particle system, we measure the enthalpies, melting points and heats of fusion of the known crystal forms as a function of aspect ratio of the molecule, and determine the various properties of the liquid state, (volumes, enthalpies, heat capacities, and ergodicity-breaking temperatures) that are needed to understand the melting point and vitrification patterns observed. A limitation of our study is that we can only obtain the melting behavior of systems whose crystal structures are known.

Fortunately, considerable information on the packing of rod and disc-like objects is already available from the granular materials community. The packing of hard oblate and prolate ellipsoids was studied by Donev et al[4,18] and it was found that packing fractions, defined by volume of the objects divided by the volume of the box containing them, rise to maxima of 0.77 for aspect ratios of $1/\sqrt{3}$ and $\sqrt{3}$, respectively. These are obtained in a structure that becomes hexagonal close-packed in the limit of spherical particles. The packing efficiency for this structure (which we call Donev-1), peaks sharply at the above values (see SI-3). Packing efficiencies for crystals of the maximum value 0.77 can be maintained at larger aspect ratios, if the structure is modified as described in ref.[3] but we do not address this range in any detail in this work. Between $\alpha = \sqrt{3}$ and $1/\sqrt{3}$, Donev et al. found a minimum packing efficiency at the hard sphere value $\alpha = 1.0$, and a very high (jammed) random packing efficiency (0.735, competitive with the crystalline form) for the fully aspherical case at $\alpha = 1.25$. Fully aspherical means that the cross section is not circular but is also ellipsoidal. (Note[4] that a single number, like $\alpha = 1.25$, can still characterize this case if the two short (lateral) dimensions are reciprocals). For the prolate ellipsoid case (lateral dimensions the same), a lower maximum for random packing efficiency was found at $\alpha =1.50$. This number will prove very relevant to our study. Related observations were made for spherocylinders[5]. These are, of course, always athermal systems, and the ordered states only spontaneously disorder during dilatation.

When interparticle attractive forces are added, as in our study, one can explore the density and temperature variables separately, and observe melting points and glass transitions etc. It is with the effect of temperature at constant pressure on such systems that the present work is concerned. By varying the aspect ratio on the one hand, and binary mixing interactions on the other, we will identify a crystal-free route to the glassy state.

For the G-B model[7] the pair interaction energy is of quite complex form (see SI-1) but, despite its complexity, the energy can still be obtained as a function of a single variable, $\alpha$ - the length-to-width (aspect) ratio of the molecules. The G-B model has three other parameters, one of which ($\alpha_2$, the ratio of energies of side-by-side to end-to-end configurations) varies with aspect ratio between the LJ extreme (where it must be unity) and the G-B model at aspect ratio 3.0 (where it has the value 5.0). However this variation can be expressed by the linear interpolation $\alpha_2 = 2\alpha -1$, so that our tuning can still be performed by variation of the single parameter $\alpha$. More complex choices can of course also be made.

In our study, we change $\alpha$ stepwise and determine the melting point of the new crystal at each step, using the method of ref.[19]. In this method, defect planes (that nucleate the liquid during heating, as efficiently as do free surfaces (see SI-2) are introduced by a rapid crystallization process with $\alpha$ not far from unity.. Each subsequent change in $\alpha$ is accompanied by a corresponding distortion of the crystal (face centered cubic FCC in the initial LJ state, and FC tetragonal, FCT in the cases of $\alpha > 1$) and gives us a new and unique system. Near and beyond $\alpha = 1.5$, the domain of stability against vibrational displacements is exceeded and the system spontaneously disorders as T rises above 0 K. At high aspect ratios, smectic B crystals become the stable state. A phase that is hexagonal close-packed in the $\alpha = 1$ limit, described by Donev et al (and called Donev-1 by us), is also found at $\alpha > 1.8$. It forms more easily than smectic B crystal during cooling at α = 2.2, and can be tuned to lower α values, but it always transforms to smectic B on heating so its melting point cannot be determined.

After each change in α (and crystal equilibration which is very rapid) the melting points of FCT crystals were determined by observing the temperature of sudden enthalpy increase, as described in ref.[19] and validated in SI-2. Results are presented graphically in SI-1, where temperatures are normalized to the G-B minimum energy (obtained for the end-to-end configuration).

The same procedure was then repeated for (defect or grain boundary-containing) crystals of the smectic B crystal type, obtained by rapid crystallization of liquids of higher α (α > 1.5).

In Figure 1a, the enthalpies of these crystal forms (both perfect, and defect-containing for melting point studies), are displayed as functions of α. Included, for comparison, are the enthalpies of glassy forms obtained by hyperquenching HQ to avoid crystallization at any aspect ratio, and also by slow cooling SC in the intermediate range (α = 1.4-1.8) where the liquids never crystallize. It is seen that only the perfect Donev-1 crystal can compete in enthalpy with the slow-cooled glass in the α = 1.4-1.6 range.

In the non-crystallizing range, the enthalpies of fusion of the crystals (produced by parameter changes at 0 K) are difficult to quantify (See SI-1 Figure S3). In the range α = 1.4-1.6, no fusion signal (jump in enthalpy) could be detected. Indeed, the crystals seemed to spontaneously disorder as thermal energy was introduced to give the lowest temperature studied (0.1). The enthalpies of fusion are shown in Figure 1b. The latter are striking: they extrapolate, from either side, to zero at the same α value, 1.45± 0.03. A vanishing of the

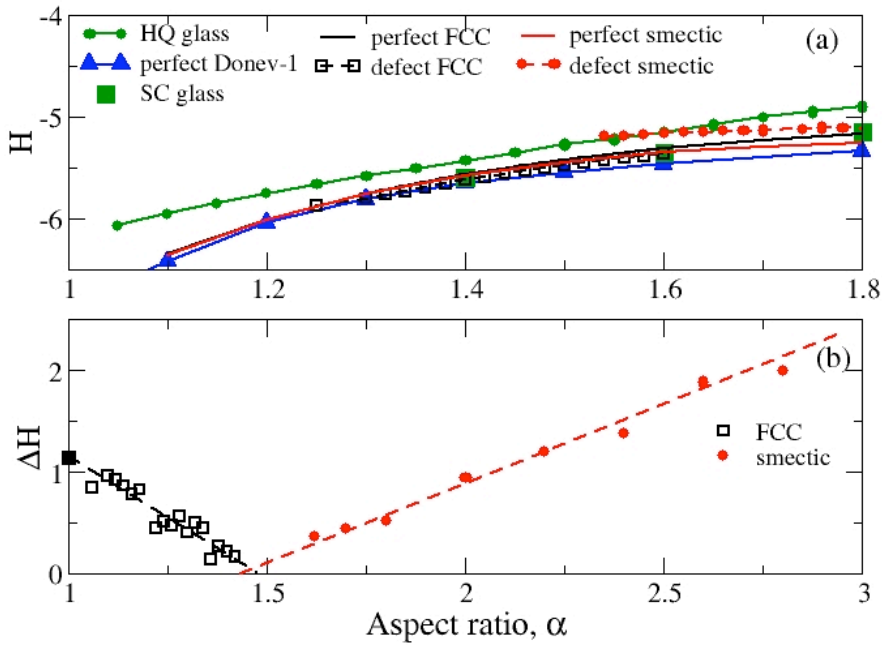

**Figure 1.** (a) Crystal enthalpies for perfect crystals FCT, Smectic B and Donev-1 at T = 0.1, compared with the enthalpies of the defect crystals obtained from freezing liquids at aspect ratios α =1.1, 3.0, and 2.2, respectively, and then tuning α within the crystal state as described in text. Included for comparison are the enthalpies of glassy states obtained by hyperquenching HQ (for all α values), and by slow cooling SC for α values in the glassforming range. Evidently, at α = 1.4, no crystal has significantly lower enthalpy than the slow-cooled glass. (b) Fusion enthalpies of crystals of different aspect ratio α, in the G-B model. Note good agreement of extrapolation to α = 1, with value for pure LJ from ref.[20]

fusion energy signature of melting has recently also been reported for a model metallic system potential[21] in which melting was observed at different fixed densities (though in this case the fusion occurs over a range of temperatures, so is not strictly first order)

We display melting points obtained from the temperatures of enthalpy discontinuities (SI-1, Figure S3) in Figure 2 and compare them with glass transition temperatures $T_g$ determined as in Figure 3.

For α near 1, the data extrapolate well to the value for pure LJ[20]. For higher α, as $T_m$ approaches $T_g$, the values will become distorted upwards because the enthalpy change used to detect melting cannot occur when the particles cannot move freely. This is because the phase generated is then not at its minimum chemical potential. All simulation-based melting point determinations will encounter difficulty when liquid states are non-ergodic (see SI-2). Superheating above the equilibrium melting point is well-known in laboratory studies of crystals with extreme viscosities at their melting points (e.g. by >165K for albite, and 450K for quartz[22]). Thus the melting points recorded in Figure 2 are excluded from approaching $T_g$ and, accordingly, the extrapolated minimum is falsely high. This explains the paradox of positive melting points in the presence of vanishing fusion enthalpies near α = 1.5.

In the range α = 1.4-1.6, no enthalpy discontinuity at all can be detected. Each of the three lattices studied is so unstable in this range that anharmonic particle oscillations broaden the radial distribution function, RDF, to glasslike values at the lowest temperatures examined (T = 0.05). During heating, only changes of slope, characteristic of glass transitions, are seen. These are discussed further below in relation to Figure 3.

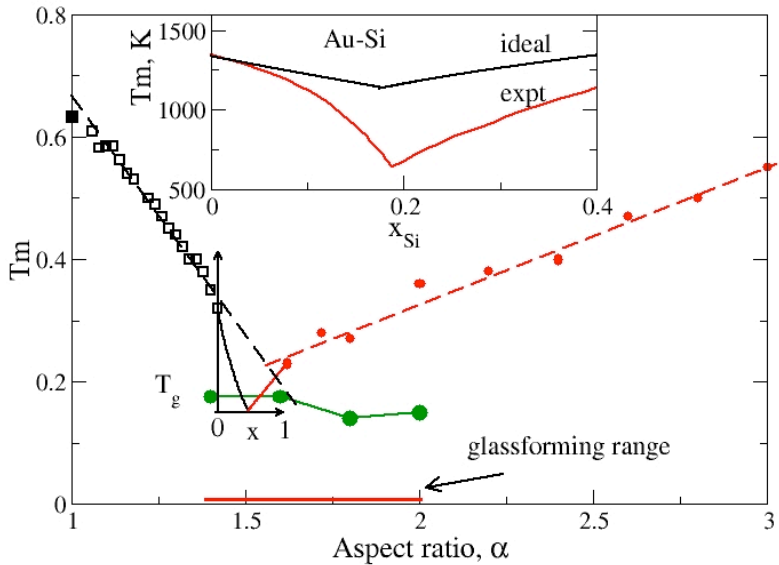

**Figure 2.** Melting points observed during heating of the defect crystals, compared with ergodicity-breaking temperatures during slow cooling, identified in Figure 3 as $T_g$.[23] Note that $T_m$ values assessed close to $T_g$ by our method (or any other) will be falsely high[22] because the equalization of chemical potentials at the melting point depends on particles being able to rearrange in space to yield the equilibrium liquid phase. **(see text).** Values at lower aspect ratio extrapolate to the ref.[20] value for pure LJ. The V-shaped addition, bridging FCT and smecticB cases with the lowest melting points, shows the liquidus temperatures for mixtures of these substances according to ideal mixing laws **(see text).** The **insert** shows the relation between the ideal mixing eutectic for the Au-Si marginal metallic glassforming system, and the actual experimental system, where the large difference is due to the optimal negative nonideal mixing effects for this system. (A binary compound forms from the glass during heating but melts/decomposes to the elements just below the eutectic temperature). With such non-ideal effects, the G-B mixture we discussed could have an extrapolated eutectic far below its $T_g$, and arguably also below the Kauzmann temperature (where $S_{liquid} = S_{crystal}$) . We consider experimental cases in a concluding discussion.

From these observations it would seem (a) none of the crystals we have studied are stable with respect to the amorphous phases near α = 1.5, and (b) even where melting can be observed, the *equilibrium* melting points of the lowest melting members of this infinite family should be located near or *below* the $T_g$. Does this mean we must identify α = 1.4 -1.5 cases as "ideal glassformers"[11]? The answer must be in the negative because we cannot prove that there are no alternative packing modes of lower symmetry – crystals that are more stable than those that we have been able to test. Intuition suggests there should be, for the single component case. However, we find the following argument in favor of the existence of ideal glassforming liquid *solutions* difficult to counter. It involves the thermodynamics of mixtures, and proceeds in two steps.

Firstly, it is well known that if two substances are insoluble in each others' crystal lattices, and mix ideally in the liquid state, then the melting point of each is lowered due to the entropy stabilization of the solution state. If the heats of fusion are known, then ideal solution laws permit the calculation of the melting point depressions and thence the eutectic temperature. Let us suppose that ellipsoids of different aspect ratios in most cases do not fit favorably on each other's crystal lattices, so that the above considerations apply to them. We can easily calculate their liquidus lines and hence the maximum liquid stability and lowest melting point for their mixtures. This is shown for mixtures of two different aspect molecules as the V-shaped addition to our "experimental" plot (Figure 2), where **x** is the mole fraction of the second component (smectic B [α = 1.6] mixed with FCT [α = 1.4]). Already the eutectic temperature is below the glass transition temperature of this study. The same considerations would apply to the hypothetical polymorph(s) of marginally higher stability, hence melting point, than those we have considered. (If there were any cases melting above 0.3, we would have observed their crystallization).

Secondly, we show how very much greater the melting point depression can be when the two components mix non-ideally with a judiciously negative heat of mixing. The insert to Figure 2 shows part of the phase diagram for the system gold-silicon[24] which yielded the first metallic glass (on splat quenching). The enthalpic drive to mix is optimal in the Au-Si case, from the point of view of lowering melting point without producing a new binary crystal of melting point higher than the two-component eutectic. The inset to Figure 2 compares the actual Au-Si liquidus lines, and eutectic temperature, with that calculated for ideal mixing using the known enthalpies of fusion. The difference is great. It shows that if, as in the Kob-Andersen model, we were to introduce attractive interactions between the G-B components that we have considered in the first step of our argument, we could obtain a much deeper eutectic than the mere ideal mixing result shown in Figure 2. It is difficult to argue that there would not, then, result a liquid that is thermodynamically incapable of crystallizing before the system $T_g$ is reached, even on laboratory cooling rates. The strength of this argument is enhanced by the next point we make concerning the nature of the liquids in the vicinity of α = 1.5.

In Figure 3, we show the enthalpies and their derivatives, the heat capacities, of G-B liquids in the range α = 1.4-2.0. Their features, in particular the hysteresis behavior, are described in the caption. Elsewhere[25], the "hysteresis peak" obtained by differencing the upscan and downscan enthalpies, has been used to define ergodicity-breaking, and the peak temperature has been shown equal to the conventional "onset heating" glass temperature[23], $T_g$. However in the domain α = 1.4-1.6, no hysteresis can be observed at all, (see Figure 3, α = 1.4), so no glass transition can be defined in this manner. On the other hand there are breaks in the *slopes* of the enthalpy vs. T plots of Figure 3a (see also Fig. S3 in SI-1) which translate to the jumps in heat capacity seen in Figure 3(b), and they are always occurring in the same temperature range, T = 0.1-0.15. Wang[26] has demonstrated that the hysteresis peaks in laboratory glassformers diminish with increasing liquid fragility, tending to vanish at the predicted high fragility limit, where $T_g$ and $T_K$ almost coincide. Indeed the detailed study of hard dumbbells by Zhang and Schweitzer[9] revealed a maximum in liquid volume fragility at a length:breadth (i.e. aspect) ratio of 1.43. Thus, at the very aspect ratio where we find no stable crystal phase, the difference between the simulation $T_g$ and the (lower) laboratory $T_g$ is minimized. In other words the case which most closely approximates the "ideal glassformer" seems also to be the case that most closely approximates the "ideal glass" ($T_g = T_K$). This observation certainly warrants further study, as the combination of extreme fragility with extreme liquid stability (also suggested recently by studies of a hard sphere dipole model[27]) would be of great intrinsic interest. We note additionally that these simple, single component, asymmetric molecular systems provide excellent models for unambiguous study of such issues as static vs. dynamic correlations (here orientational) and their characteristic lengths[28-32] and should therefore be very useful in resolving some of the more contentious issues that currently vex the "glasses" community.

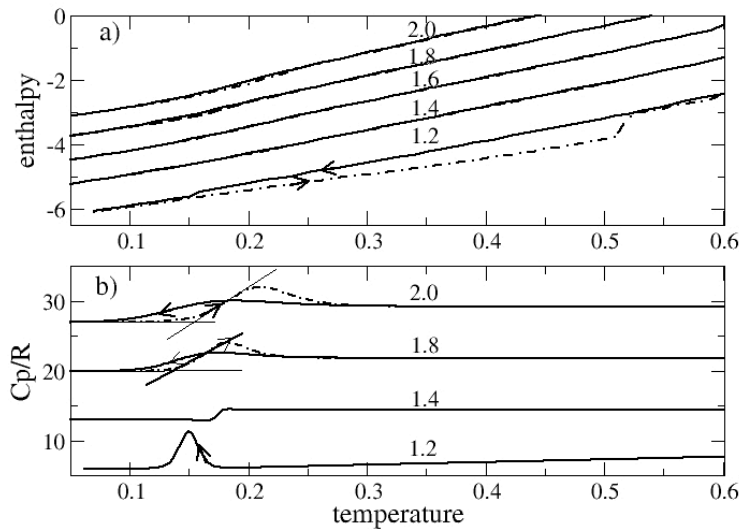

**Figure 3.** (a) enthalpies of the liquid and glassy states of the G-B model, for five values of aspect ratio α, spanning the pseudo triple point in the temperature-potential phase diagram (Figure 2), and (b) the corresponding heat capacities, Enthalpies are shifted by 0.5 units from each other for clarity. The $C_p$ plots are separated by 7 units from each other (low temperature value is the classical 3R in each case). The jump in $C_p$ at $T_g$ is 33% of the glass value in each of the glassformer cases shown (α = 1.4, 1.8, and 2.0). At α = 1.4 and 1.6 (not shown) the normal enthalpy hysteresis of glass-formers disappears and the liquid behaves like the maximally fragile liquid case of Wang's hysteresis analysis[26]. The behavior at α = 1.2, where crystals nucleate at the $T_g$ of the other liquids, is surprising (the subsequent crystal melting point accords with Fig. 2 values).

      To conclude, we return to experiments to make a simple observation, based on the insights gained from this work. It seems not to have been made before, and it almost guarantees the existence of crystal-free routes to the glassy state. We apply the observation that optimum non-ideal mixing thermodynamics can lower melting points to 0.5 of their pure substance value (Fig. 2 insert), to the mixing of two pure substances that

individually obey the "2/3 rule" ($T_g = 2/3 T_m$). The eutectic temperature could then lie as low as 0.5 of $T_m$ which is clearly far below the $T_g$ of either component. And since the mixing is non-ideal negative, the solution will be ordered by the interaction, so that the solution $T_g$ s will lie above the component values, as found in e.g. Lesikar's studies of alcohol + Lewis base mixtures[33]. Thus not only would the equilibrium fusion temperature ($T_{eutectic}$) lie well below $T_g$ but it would also lie below $T_K$, which is typically found at 0.8 $T_g$. Such solutions would clearly be ideal glassformers. In the pharmaceutical literature there are reported many weak acids and weak base glassformers with $T_g > 2/3\ T_m$, some[12] as high as $0.8T_m$. By choice of $pK_a$ difference, their mixtures could be tuned to optimize the non-ideality of mixing, thus many ideal glassformers should be available.

**Methods.** The standard molecular dynamics methods used, the Gay-Berne potential, and the melting point characterization validation, are fully described in SI-1 and 2.